# Separating Electrons and Donors in BaSnO$_3$ via Band Engineering

Abhinav Prakash[1,†,*], Nicholas F. Quackenbush[2,†,*], Hwanhui Yun[1], Jacob Held[1], Tianqi Wang[1], Tristan Truttmann[1], James M. Ablett,[3] Conan Weiland,[2] Tien-Lin Lee,[4] Joseph C. Woicik,[2] K. Andre Mkhoyan[1], and Bharat Jalan[1,*]

[1]Department of Chemical Engineering and Materials Science, University of Minnesota, Minneapolis, MN 55414, USA

[2]Materials Measurement Science Division, Material Measurement Laboratory, National Institute of Standards and Technology, Gaithersburg, MD 20899, USA

[3]Synchrotron SOLEIL, L'Orme des Merisiers, Boîte Postale 48, St Aubin, 91192 Gif sur Yvette, France

[4]Diamond Light Source, Ltd., Harwell Science and Innovation Campus, Didcot, Oxfordshire OX11 0DE, England, UK

[†]Equal contributing authors

[*]Corresponding authors: Abhinav Prakash (praka019@umn.edu), Nicholas F. Quackenbush (nicholas.quackenbush@nist.gov), Bharat Jalan (bjalan@umn.edu)

**Abstract**

Through a combination of thin film growth, hard X-ray photoelectron spectroscopy (HAXPES), magneto-transport measurements, and transport modeling, we report on the demonstration of modulation-doping of $BaSnO_3$ (BSO) using a wider bandgap La-doped $SrSnO_3$ (LSSO) layer. Hard X-ray photoelectron spectroscopy (HAXPES) revealed a valence band offset of $0.71 \pm 0.02$ eV between LSSO and BSO resulting in a favorable conduction band offset for remote doping of BSO using LSSO. Nonlinear Hall effect of LSSO/BSO heterostructure confirmed two-channel conduction owing to electron transfer from LSSO to BSO and remained in good agreement with the results of self-consistent solution to one-dimensional Poisson and Schrödinger equations. Angle-dependent HAXPES measurements revealed a spatial distribution of electrons over 2-3 unit cells in BSO. These results bring perovskite oxides a step closer to room-temperature oxide electronics by establishing modulation-doping approaches in non-$SrTiO_3$-based oxide heterostructure.

The realization of two-dimensional electron gases (2DEGs) is one of the most celebrated discoveries in compound semiconductors owing to their high mobility and quantum confinement effects.[1,2] Historically, one of the ways of achieving 2DEGs in compound semiconductors is via modulation doping approach where a semiconductor *A* can be remotely doped by interfacing it with semiconductor *B* that has a higher conduction band minimum. The conduction band offset ensures the transfer of electrons from *B* to *A*. In this scheme, the source of electrons is the chemical dopants in semiconductor *B* and therefore, they are also tunable. The benefits of modulation doping stem largely from the fact that they are spatially separated from their donor ions. Modulation-doped structures in compound semiconductors have yielded 2DEGs with exceptionally high mobilities both at low temperature and room temperature (RT).[1,2] In comparison, 2DEGs in complex oxides are relatively less explored.[3-5] The 2DEGs in complex oxides are typically created *via* polar discontinuity mechanism using the polar and non-polar interfaces such as $LaAlO_3/SrTiO_3$ and $RTiO_3/SrTiO_3$ (where *R* is a rare-earth element).[4,6-8] Although, these structures have yielded high-mobility 2DEGs at low temperatures exceeding 50,000 $cm^2/Vs$,[9] their RT mobility has remained below 10 $cm^2/Vs$.[10-12] This is mainly due to the strong electron-phonon coupling in $SrTiO_3$.[12,13] Although other factors such as interface roughness, dislocations, and point defects can also affect 2DEGs' mobilities, electron mobilities at RT are largely limited by strong electron-phonon scattering in $SrTiO_3$. Attempts to synthesize non-$SrTiO_3$ based modulation-doped heterostructure have been unsuccessful so far despite theoretical predictions.[14,15]

Recently, alkaline earth stannates with perovskite structure have emerged as promising candidates for high RT mobility and high optical transparency.[16-22] High RT electron mobility is attributed to the low electron effective mass and weak electron-phonon coupling.[15,19] $SrSnO_3$ (SSO) possesses wider bandgap (4 – 4.5 eV) and can be doped *n*-type in thin films with reasonably high electron mobility, 55 cm$^2$/Vs at RT.[23, 24] Additionally, SSO films can be grown coherently.[24] BaSnO$_3$ (BSO), on the other hand, possesses lower bandgap of ~3 eV,[15, 25] a relatively larger lattice parameter (4.116 Å), and has not yet been grown fully strained on a commercially available substrate. The latter introduces misfit and threading dislocations in BSO films limiting electron mobilities below that of bulk single crystals.[18-21, 26] While some progress has been made in addressing dislocation issues using undoped buffer layer [18-21, 26, 27] or by developing lattice-match substrates [28], very little has been done to eliminate scattering due to ionized donors.

In this paper, we report on the demonstration of a modulation doping approach separating electrons and charged dopant ions in BSO using a band-engineered heterostructure grown by hybrid molecular beam epitaxy (MBE). Using hard X-ray photoelectron spectroscopy (HAXPES), scanning transmission electron microscopy (STEM), and electrical transport, we establish the band alignment, location of electron gas, and their spatial distribution in La-doped SSO (LSSO)/BSO heterostructures in addition to demonstrating insulator-to-metal transition in LSSO owing to an internal charge transfer.

We first present the structural data of the heterostructure investigated in this work. Figure 1a shows the schematic of the heterostructure consisting of GdScO$_3$ (GSO) (110)/ 46 nm BSO/ 14 nm SSO/ 25 nm BSO/ 1 nm SSO/14 nm LSSO. A 46 nm-thick BSO buffer layer was grown on GSO to obtain relaxed, insulating BSO film as a template for subsequent film growth. A fully-strained 14 nm SSO layer was then grown followed by 25 nm BSO layer in an attempt to constrain threading dislocations in the bottom BSO buffer layer. An undoped 1 nm SSO was used between BSO and LSSO as a spacer layer to provide a larger spatial separation between charge carriers and donor ions for modulation doping. For brevity, we will refer this structure as LSSO/BSO heterostructure. Figure 1b shows a wide-angle X-ray diffraction (WAXRD) scan of this structure showing (002) film peaks, (002) substrate peak and thickness fringes. Analysis of the XRD data yielded an out-of-plane lattice parameter of 4.131 Å ± 0.002 Å for BSO layers (mostly relaxed), 4.010 Å ± 0.002 Å for SSO layers consistent with a fully strained SSO film on BSO, and layer thicknesses in excellent agreement with their intended structure as shown in the figure 1a. Grazing incidence X-ray reflectivity (GIXR) scan with well-defined Kiessig fringes (figure 1c) further confirms uniform film thicknesses consistent with WAXRD data. GenX fits also yielded interface roughness for each interface, < 1-2 unit cell (u.c.), which is consistent with our STEM analysis, which combines high-angle annular dark-field (HAADF) imaging with the energy-dispersive x-rays spectroscopy (EDX). The EDX elemental map of Ba $L_\alpha$ (green) and Sr $L_\alpha$ (blue) of the SSO/BSO heterostructure, presented in figure 1d shows uniform film thickness and absence of any phase segregation. In addition, atomic-resolution HAADF STEM image of the top 14 nm LSSO/1 nm SSO/ 25 nm BSO interface indicates a reasonably smooth interface (figure 1e) and no presence of misfit dislocations in SSO. However, inspection of STEM images shows a large number of threading dislocations in BSO due to strain relaxation.

We now turn to the discussion of electrical transport data. Figure 2a shows sheet resistance ($R_s$) vs. $T$ plot for LSSO/BSO heterostructures consisting of 14 nm (red solid line) and 7 nm (green solid line) LSSO layers indicating insulating behaviors with $R_s > h/e^2$ at all temperatures. For reference, we also show $R_s$ vs. $T$ as an inset for a representative 12 nm LSSO film without a BSO interface layer revealing a metallic behavior with significantly lower sheet resistance. For doping, the La cell temperature was kept fixed at 1200 °C. The observed behavior is significantly different for LSSO when it is interfaced with BSO. This result suggests that electrons are either trapped in structural-related defects in SSO or may have transferred to BSO layer with low mobility accompanied by a metal-to-insulator transition in LSSO. As discussed above, no noticeable structural defects were observed in LSSO film grown on BSO. Rather, our Hall measurements showed nonlinear behavior associated with two-channel conduction.

Figure 2b and 2c show transverse resistance, $R_{xy}$ at 30 K as a function of $B$ revealing a linear Hall slope for the 14 nm LSSO/BSO heterostructure and a nonlinear Hall slope for the sample with 7 nm LSSO layer. Note the dopant density was kept identical in these samples. While the linear Hall slope does not explicitly rule out two-channel conduction, the nonlinear behavior in 7 nm LSSO/BSO sample clearly suggests the presence of two-channel conduction. The fitting of non-linear Hall behavior at low $B$-field and high $B$-field yielded a nominal carrier density of $7.24 \times 10^{12}$ and $1.16 \times 10^{13}$ cm$^{-2}$ respectively. The low magnetic field density typically results from high-mobility carriers whereas at high-fields, all carriers can contribute to the Hall signal. Two-channel conduction model didn't produce reliable fits to the data given there are four variables. For this reason, we don't report on the results of fitting. Rather, we estimated carrier density using low-field slope and high-field slope suggesting *qualitatively* electron density in the order of mid $10^{12}$ cm$^{-2}$ in the two channels, which is near or below the critical density for metal-to-insulator transition in stannates as reported earlier.[19, 29, 30] We argue that it is this redistribution of electrons across the interface, which makes LSSO/BSO insulating whereas LSSO without BSO remains metallic. Additionally, interfacial scattering and scattering from threading dislocations in BSO can also play important roles in localizing the carriers. Significantly however in agreement with the nonlinear Hall data, HAXPES measurements revealed the electron transfer from LSSO to BSO owing to modulation doping due to a straddling type I band alignment, as discussed below.

To investigate the band alignment, we first measure the valence band (VB) photoemission of the two reference materials as shown in figure 3a. The valence band maxima ($E_V$) are determined from the linear fit to the leading edge of the main valence band and extrapolating it to zero intensity.[31] The small density of states observed between 2 and 3 eV above the main edge are due to growth-related defects, e.g., dislocations and/or point defects. The valence band maxima were found to be 3.99 eV ± 0.02 eV and 3.18 eV ± 0.02 eV for LSSO and BSO, respectively. In addition to the valence band, the LSSO HAXPES spectrum also displays a weak feature with a sharp edge at the Fermi level (magnified 100×), representing occupied Sn 5s states at the bottom of the conduction band. This is confirmed to be correlated to the presence of La in the core level HAXPES spectrum (see inset) and is analogous to the well-studied La-doping in BSO, due to their similar electronic structure.[32-34] This provides a reasonable estimate of the bandgap of our LSSO, since the conduction band minimum ($E_C$) is nearly degenerate with $E_F$, thus $E_V \approx E_G$. Undoped BSO however, does not have this conduction band filling and so $E_V$ can only be considered as a lower limit of the band gap. Our value of 3.18 eV is in good agreement with previous reported thin film samples.[25, 35, 36] The valence band HAXPES spectrum of the LSSO/BSO heterostructure is shown in figure 3b. Due to the inherent surface sensitivity of photoemission, the VB spectrum is dominated by the top LSSO layer, however the high kinetic energy of HAXPES allows the Ba 5p doublet near 15 eV from the buried BSO layer to be observed. To determine the valence band offset at this buried interface, the valence band spectrum of the heterostructure is fitted as a linear combination of the spectra collected from each reference material, allowing the binding energy alignment to be determined by the fit. The resulting components of the fit are displayed in figure 3b with dotted lines. We thus find the VB offset to be 0.71 ± 0.02 eV. Figure 3c shows the energy level flat-band diagram for the heterostructure based on our HAXPES measurements indicating a conduction band offset of +0.10 eV ± 0.02 eV between LSSO and BSO.

To get further insights into the transport data discussed in figure 2, we calculated the band diagram using experimental band offsets for the LSSO/BSO heterostructure (figure 3d). The band diagram was calculated using 1D Poisson solver.[37] For calculation, dielectric constants of 20 and 17 were used for SSO and BSO, respectively.[38] La dopant density of $8.5 \times 10^{19}$ cm$^{-3}$ ($n_{2D} = 1.2 \times 10^{14}$ cm$^{-2}$) in the 14 nm LSSO layer were used. Figure 3d (top panel) reveals the presence of lower potential region (shaded in green) for electrons on the BSO side of the interface in addition to confirming that a fraction of electrons from LSSO can transfer towards the BSO side. Figure 2d (bottom panel) shows the calculated 3D electron density profile across LSSO/BSO as a function of depth yielding an expected carrier density of $9 \times 10^{13}$ cm$^{-2}$ and $5 \times 10^{12}$ cm$^{-2}$ on the LSSO and BSO respectively, as a result of modulation doping and electron transfer. Our result is in good agreement with the DFT calculations.[15]

We look again to HAXPES to investigate the location and spatial distribution of the conduction electrons at $E_F$. By resolving the emission angle of the ejected photoelectrons, a depth profile can be achieved. Figure 4a shows the angle-integrated (traditional) VB HAXPES spectrum (top) along with the emission-angle resolved 2D spectrum (bottom). The 2D spectrum was analyzed by dividing into 5 angular ranges (centered at 82°, 71°, 61°, 51°, 40°) and summing to create 5 VB spectra with varying depth sensitivity. Figures 4c-e show the extracted shallow core level regions, Sr 4$p$ and Ba 5$p$, after background removal, as well as the region near $E_F$, where a small density of states is observed as in the reference LSSO. The intensity of these $E_F$ states decreases with decreasing emission angle, i.e., as the measurement becomes more surface sensitive, indicating that they do not simply reside in the top LSSO layer. In fact, the intensity profile as a function of angle, shown in figure 4b, exhibits a profile more similar to the Ba 5*p* of the buried BSO layer than the Sr 4*p* of the layer.

To extract more quantitative depth information, these normalized intensity profiles are modeled based on the exponential attenuation of the escaping photoelectrons.[39, 40] The intensity of photoelectrons measured at the analyzer is $I = I_0*\exp[-t/(\lambda \sin \alpha)]$, where, $\alpha$ is the emission angle, $t$ is the thickness of the overlayer the photoelectrons must traverse, and $\lambda$ is the effective attenuation length, which can be calculated. Here, $\lambda$ was calculated to be 7.7 nm for SSO at the photon energy and polarization geometry used via the TTP-2M equation[41, 42] and accounting for the single scattering albedo.[43] The Sr 4*p* and Ba 5*p* profiles fit well when modeled as arising from the top 15 nm or buried under such a layer, respectively. Following this same analysis for the $E_F$ states, and allowing the interface thickness to be a fit parameter gives the best fit when a finite intensity of these $E_F$ states arise from the SSO layer, while the majority are from deeper than 15 nm, i.e. the top layers of the BSO. The fit reveals a thickness of this interface layer to be about 1.5 nm ± 0.5 nm (3 - 4 u.c.).

Therefore, it is evident that electrons transfer from LSSO to BSO owing to straddling straddling type I band alignment and that these electrons are spread over 3 - 4 u.c. in BSO. However, one may still argue La interdiffusion from LSSO to BSO to be a source of electrons in BSO buried layer. To eliminate the possibility of interdiffusion, we performed atomic-resolution STEM/EELS analysis of the LSSO/SSO/BSO interface. Figure 5a shows the atomic-resolution annular dark field (ADF) STEM image of the interface and the positions from where EELS spectra were acquired. Spectra of O *K* edge, and Ba and La $M_{4,5}$ edges collected across the interface are shown in figure 4c and 4d respectively. Due to weak EELS signal for Sr, we used fine structure of the O *K* edge to analyze Sr distribution across the interface. O *K* spectra marked as bold solid lines in figure 4c were used as a reference for *bulk* BSO and SSO (regions away from the interface). Figure 4b shows distribution of Sr and Ba across the interface using linear superposition of O *K* edge spectra obtained from *bulk* SSO and BSO. Figure 4b also depicts Ba- and La-profile across the interface determined from the analysis of EELS data shown in figure 4d. It is noteworthy that Ba-distribution determined from EELS $M_{4,5}$ edge and O *K* edge follows nearly identical trend proving viability of method for determining Sr distribution across the interface. These results show no measurable La concentration in the BSO buried layer providing strong evidence against the interdiffusion as a source of electron in BSO. The La curve follows an error function similar to the O *K* (SSO) profile, albeit shifted ~ 1 nm away from BSO, due to the presence of SSO spacer layer.

In summary, we have demonstrated modulation doping in LSSO/BSO heterostructures revealing a straddling type I band offset. Using electrical transport and HAXPES, the transfer of electrons from LSSO to BSO is confirmed which was accompanied by the metal-to-insulator transition in LSSO due to charge distribution. Angle-resolved HAXPES yielded a thickness of 3 - 4 u.c. over which electrons are spread in BSO. Although we showed electrons were separated from the donor ions, transport in BSO remains limited by threading dislocations and weaker confinement. We argue that LSSO/BSO can provide an ideal model material system for realizing high-mobility 2DEGs in complex oxides at RT if the band offsets can be increased either through alloying or strain tuning.

**Methods:**

1. <u>Hybrid molecular beam epitaxy of SSO/BSO heterostructures</u>

   SSO/BSO heterostructures were grown using hybrid molecular beam epitaxy. This approach employs a chemical precursor – hexamethylditin (HMDT) for Sn, conventional solid sources for Sr, Ba, and La (ultra-high purity of > 99.99%), and an RF plasma for oxygen. La was used as an *n*-type dopant for the doped SSO layer. Films were grown using co-deposition in an ultra-high vacuum MBE chamber (EVO-50, Omicron) with a base pressure of $10^{-10}$ torr. Beam equivalent pressures of $5\times10^{-8}$ torr and $2.5\times10^{-6}$ torr were used for Ba and HMDT, respectively for the growth of stoichiometric BSO, whereas BEPs of $3\times10^{-8}$ torr and $1.0\times10^{-6}$ torr were used for Sr and HMDT for growth of stoichiometric SSO films. La cell temperature was maintained at 1200 °C during the growth of La-doped SSO layer. An oxygen pressure of $5\times10^{-6}$ torr was used. The plasma was operated at 250 W with deflection plates kept at 250 V preventing high-energy oxygen ions from reaching the growth surface.

2. <u>Scanning Transmission Electron Microscopy sample preparation and imaging</u>

   Cross-sectional transmission electron microscopy sample was prepared by using FEI Helios Nanolab G4 dual-beam focused ion beam (FIB). The samples were thinned using a 30 kV Ga-ion beam and further polished using a 2 kV Ga-ion beam to minimize FIB-induced damage at the surface. Scanning transmission electron microscopy (STEM) experiments were performed using an aberration-corrected FEI Titan G2 60-300 STEM equipped with a CEOS DCOR probe corrector, super-X system for energy dispersive X-ray (EDX) spectroscopy, and a monochromated and a Gatan Enfinium ER spectrometer for electron energy-loss spectroscopy (EELS). Annular dark-field STEM images and EDX elemental maps were acquired at 200 kV with a beam current of ~40 pA, where the semi-convergent angle of the probe was 25 mrad. The inner ADF detector angles were 55 mrad and 96 mrad for ADF and HAADF imaging, respectively. Monochromated STEM-EELS measurements were carried out at 200 keV with screen current of ~25 pA, where probe semi-convergent angle was 17 mrad and the EELS collection angle was 29 mrad. Dual EELS mode was used to acquire low-loss, including zero-loss peak (ZLP), and high-loss EELS spectra, simultaneously. Energy dispersion of 0.1 eV per channel was used and the energy resolution was 0.4 eV.

3. Electronic transport measurements and simulation of band alignment

   Electronic transport measurements were performed in a Quantum Design Physical Property Measurement System (PPMS Dynacool) to extract the carrier density, sheet resistance, and carrier mobility. Indium was used as an Ohmic contact. Measurements were taken at temperatures between 2 K and 300 K and the magnetic field range was -9 T to +9 T. The band alignment between SSO and BSO were simulated using 1D Poisson solver by Gregory Snider, which solves the Schrodinger and Poisson equation self-consistently.[37] A 0.1 eV Schottky barrier was assumed at the La-doped SSO surface for calculating the band profile.

4. Hard energy x-ray photoelectron spectroscopy

   Hard X-ray photoelectron spectroscopy (HAXPES) was performed at beamline I-09 at Diamond Light Source (UK) with 5.930 keV photon energy using a Si (111) double crystal monochromator followed by a Si (004) channel-cut high-resolution monochromator. The hemispherical photoelectron analyzer was set to 200 eV pass energy resulting in an overall experimental resolution of 200 meV as determined from fitting a Fermi function to the valence band of a reference gold foil. The binding energy axis was calibrated using the Fermi level and Au 4*f* core lines of the gold foil in electrical contact with the sample. The X-rays were 10° glancing incidence on the sample surface and the cone of the photoelectron analyzer was oriented parallel to the polarization vector of the incident X-ray beam. Angle-resolved valence band HAXPES was performed in a fixed geometry using a wide-angle lensing mode for parallel detection of photoelectrons over a range of ~ 56° with the x-ray incidence angle fixed at 30°.

**Data availability:**

The data that support the main findings of this study are available from the corresponding authors on request.

**Acknowledgements**

The authors thank C. J. Powell for discussions regarding effective attenuation length calculations. This work was primarily supported through the Young Investigator Program of the Air Force Office of Scientific Research (AFOSR) through Grant No. FA9550-16-1-0205. Part of this work is supported by the National Science Foundation through DMR-1741801, and partially by the UMN MRSEC program under Award No. DMR-1420013. Parts of this work were carried out in the Characterization Facility, University of Minnesota, which receives partial support from NSF through the MRSEC program. Portions of this work were conducted in the Minnesota Nano Center, which is supported by the National Science Foundation through the National Nano Coordinated Infrastructure Network (NNCI) under Award Number ECCS-1542202. A.P. acknowledges support from University of Minnesota Doctoral Dissertation Fellowship. Parts of this research were performed while N.F.Q. held a National Institute of Standards and Technology (NIST) National Research Council (NRC) Research Postdoctoral Associateship Award at the Material Measurement Lab. We thank Diamond Light Source for access to beamline I-09 (SI15845-1) that contributed to the results presented here.

**Figures (Color Online):**

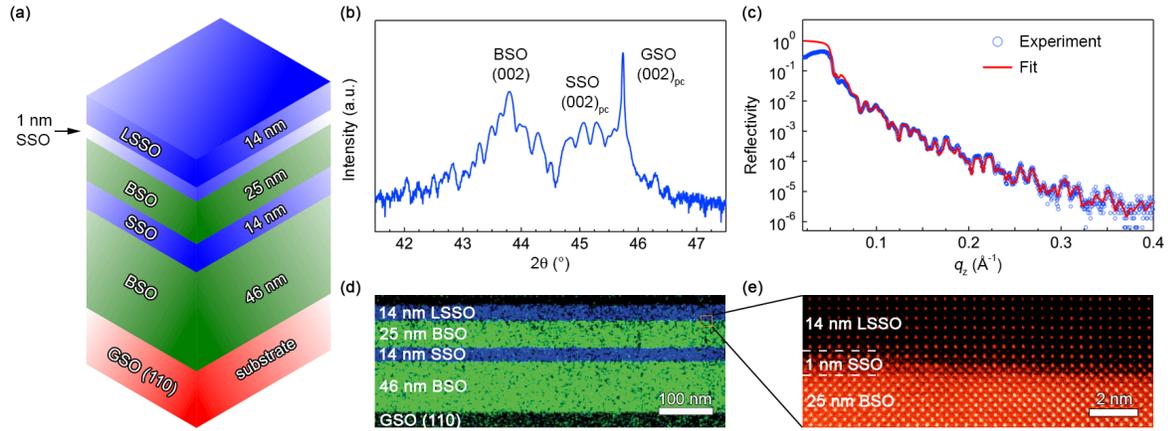

Figure 1: **Structural characterization of SrSnO$_3$/BaSnO$_3$ heterostructure.** (a) Schematic of the SrSnO$_3$/BaSnO$_3$ heterostructure grown on GSO (110), (b) Wide-angle x-ray diffraction (WAXRD), (c) Grazing incidence x-ray reflectivity of the heterostructure along with a fit using GenX software, (d) Energy dispersive x-ray (EDX) elemental maps of Ba $L_\alpha$ (green) and Sr $L_\alpha$ (blue) in the heterostructure, (e) HAADF-STEM image of the top LSSO/SSO/BSO interface.

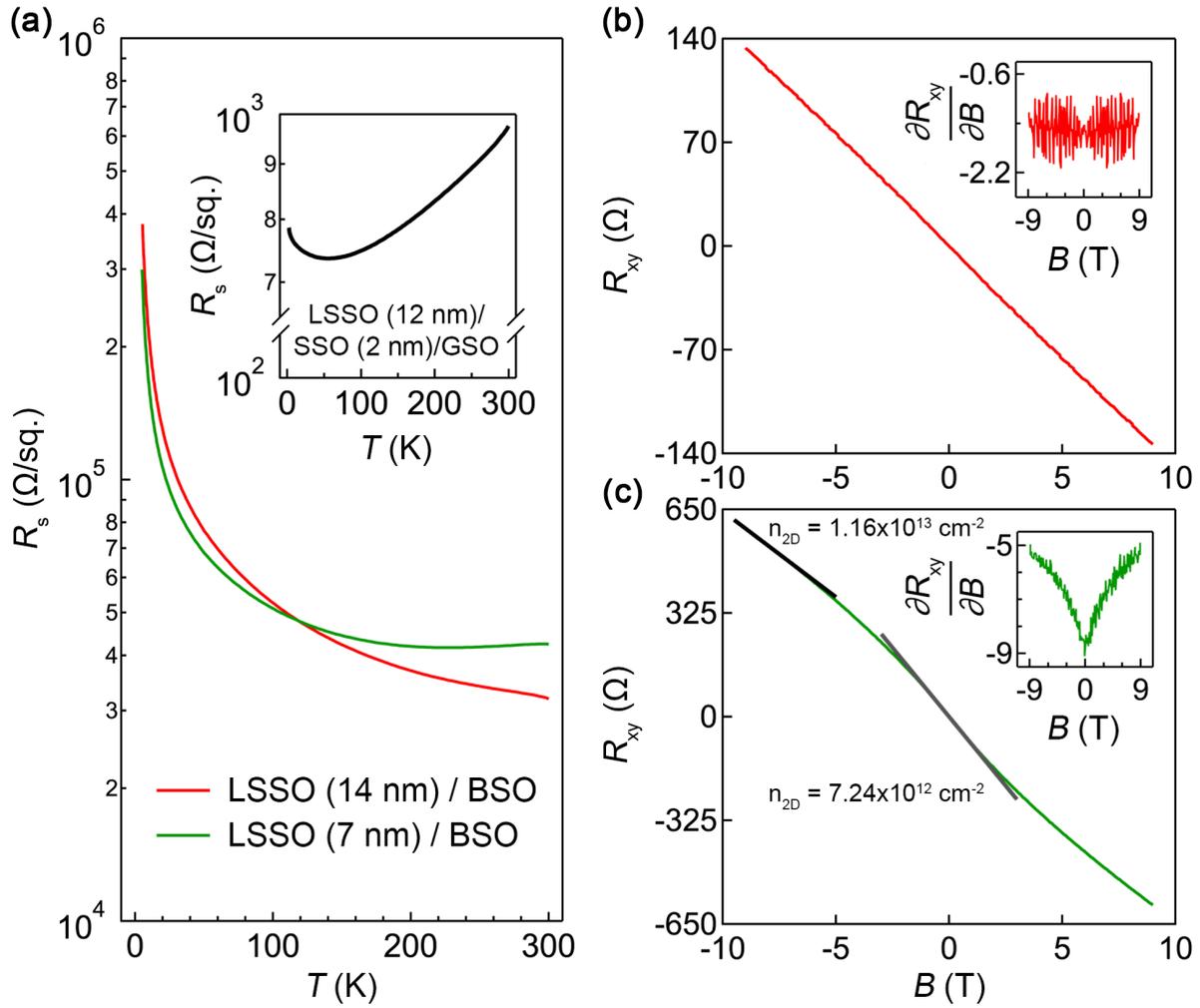

Figure 2: **Electrical Transport in SrSnO₃/BaSnO₃ heterostructures.** (a) $R_s$ vs. $T$ for the LSSO/BSO heterostructures with −14 nm LSSO (red) and 7 nm LSSO (green). Inset shows the $R_s$ vs. $T$ behavior for 12 nm LSSO/2 nm SSO/ GSO (110) without BSO interface layer as a reference, (b, c) Transverse resistance ($R_{xy}$) as a function of magnetic field ($B$) at 30 K for the two heterostructures. Insets show the corresponding first derivatives of $R_{xy}$ with respect to $B$ ($\partial R_{xy}/\partial B$) vs. $B$.

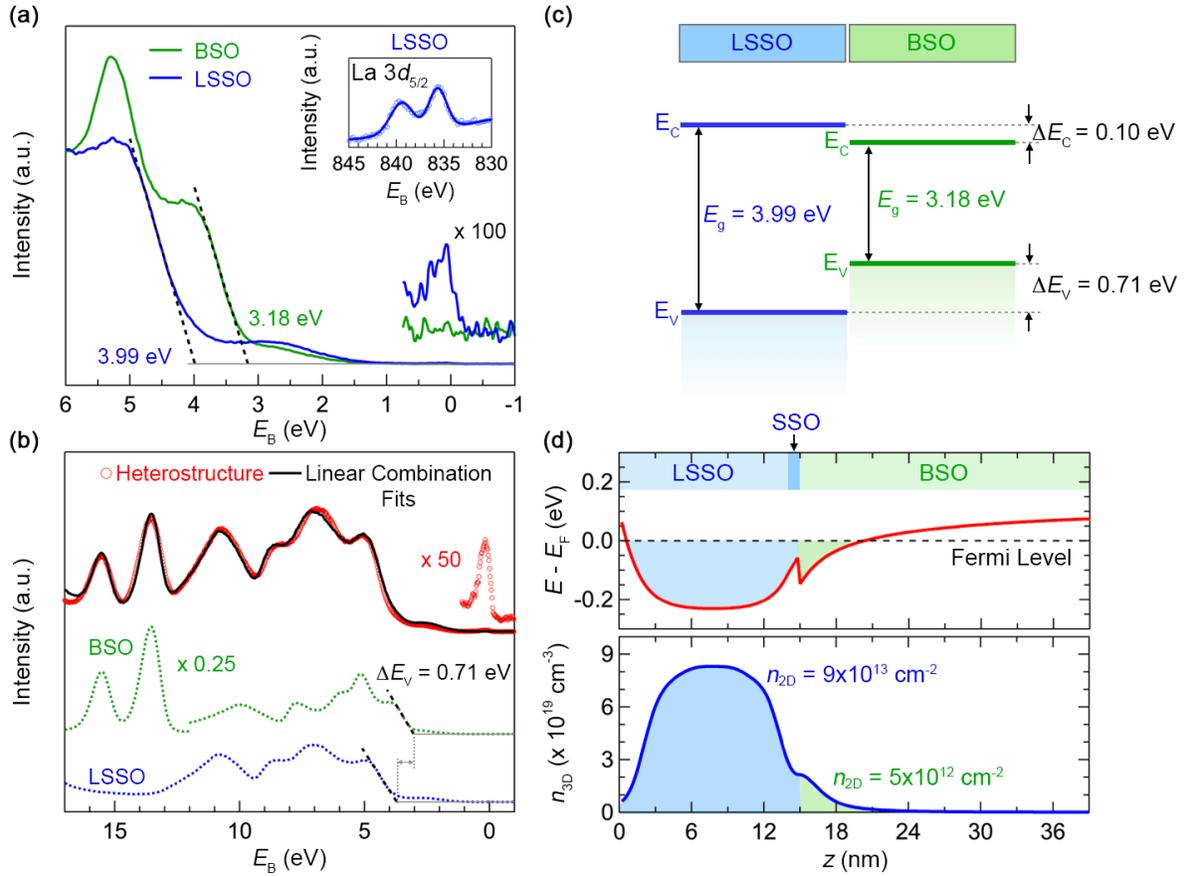

Figure 3: **Band alignment at LSSO/SSO/BSO interface.** (a) Valence band spectra of the reference BSO (green) and LSSO (blue) films. Electronic states near the Fermi states are magnified. Inset shows the La $3d_{5/2}$ core-level x-ray photoelectron spectra, (b) Valence band spectra of the SSO/BSO heterostructure (red) along with the fit (black) using linear combination of the reference valence band spectra (dotted green and blue lines) to determine the valence band offset. (c) Energy-level flat-band diagram showing the measured band offsets between LSSO and BSO, and (d) Conduction band minima (red) referenced to the Fermi level (top panel) and 3D carrier density, $n_{3D}$ (blue) as a function of depth for the SSO/BSO. The shaded regions indicate 2D density in LSSO and BSO layers after the charge transfer.

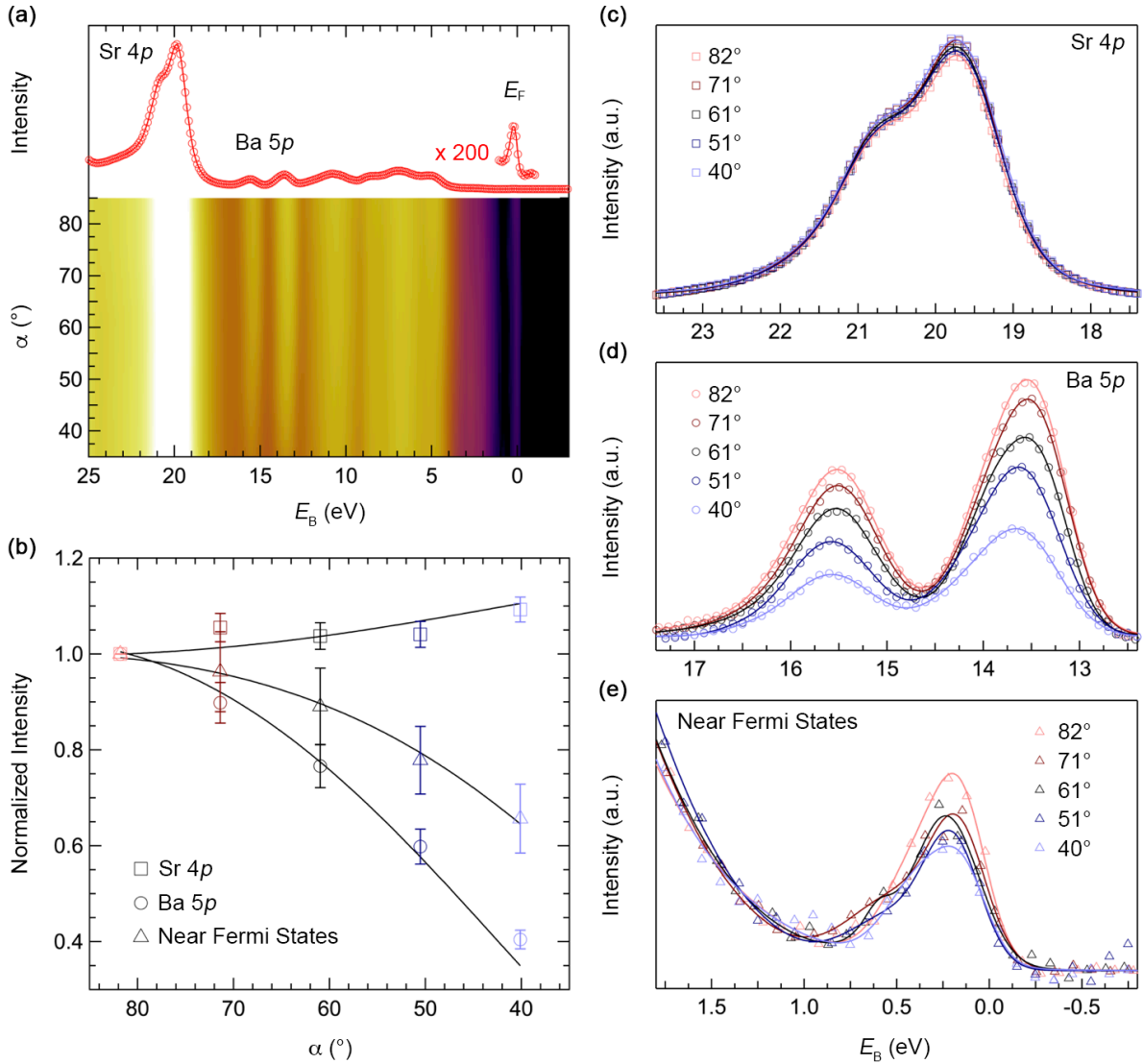

Figure 4: **Angle-dependent x-ray photoemission.** (a) Angle-resolved HAXPES valence band spectrum of LSSO/BSO heterostructure. Integrated intensity over all emission angles is shown in the top panel, (b) Normalized intensity as a function of emission angle for Sr 4*p* (squares), Ba 5*p* (circles) core levels and near Fermi states (triangles), (c-e) X-ray photoemission spectra of Sr 4*p*, Ba 5*p*, and near Fermi states.

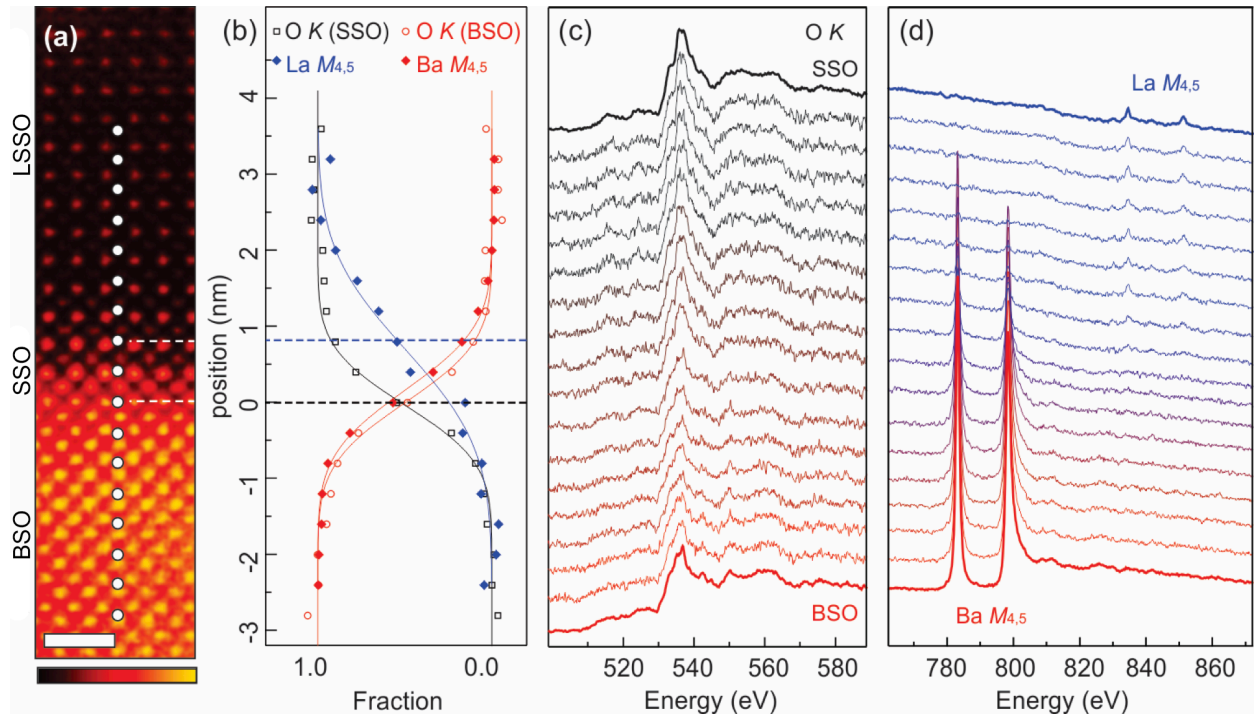

Figure 5: **Core-loss STEM-imaging and EELS obtained at the LSSO/SSO/BSO interface.** (a) Atomic-resolution ADF-STEM image of the LSSO/SSO/BSO interface. The white solid circles mark the position where EELS data were acquired. Scale bar is 1 nm. (b) The fraction of each element estimated from core-loss EELS across the interface. EELS spectra from across the interfaces for O $K$ edge (c), and Ba and La $M_{4,5}$ edges (d). The results in panel (b) were fitted to a standard error function, and the mean positions, $(x_0)$ of the erf$(x - x_0)$ are marked with dashed lines.